\documentclass[12pt]{article}

\usepackage{amsmath,amssymb,graphicx,multicol} 
\usepackage{epsf,color}
\usepackage{pstricks}
\usepackage{cite}

\newcommand{\beq}{\begin{eqnarray}}
\newcommand{\eeq}{\end{eqnarray}}

\newcommand{\centeron}[2]{{\setbox0=\hbox{#1}\setbox1=\hbox{#2}\ifdim
                                        
\wd1>\wd0\kern.5\wd1\kern-.5\wd0\fi
\copy0

\kern-.5\wd0\kern-.5\wd1\copy1\ifdim\wd0>\wd1
                                       \kern.5\wd0\kern-.5\wd1\fi}}
\newcommand{\ltap}{\>\centeron{\raise.35ex\hbox{$<$}}
                               {\lower.65ex\hbox{$\sim$}}\>}
\newcommand{\gtap}{\>\centeron{\raise.35ex\hbox{$>$}}
                               {\lower.65ex\hbox{$\sim$}}\>}

\newcommand\ZZ{\hbox{\zfont Z\kern-.4emZ}}
\font\zfont = cmss10 

\newcommand{\eq}[1]{(\ref{eq:#1})}

\begin{document}
\begin{titlepage}
\begin{flushright}
\end{flushright}

\vskip.5cm
\begin{center}
{\huge \bf 
Colored Unparticles}

\vskip.1cm
\end{center}
\vskip0.2cm

\begin{center}
{\bf
{Giacomo Cacciapaglia}, {Guido Marandella}, {\rm and}
{John Terning}}
\end{center}
\vskip 8pt

\begin{center}
{\it
Department of Physics, University of California, Davis, CA  
95616} \\
\vspace*{.5cm}
\vspace*{0.3cm}
{\tt  cacciapa@physics.ucdavis.edu, maran@physics.ucdavis.edu, terning@physics.ucdavis.edu}
\end{center}

\vglue 0.3truecm

\begin{abstract}
\vskip 3pt
\noindent
We examine a scenario where the new physics at the LHC includes an approximate conformal field theory, where some of the degrees of freedom (aka ``unparticles") carry a color charge. We present a simple argument showing that the production cross section for scalar unparticles mediated by a gauge interaction is given by $2-d$ times the standard particle expression, where $d$ is the scaling dimension of the unparticle field. We explicitly check that this is indeed the case, which involves non trivial cancellations between different Feynman diagrams, for the process $q \bar q \to$ scalar unparticles.

\end{abstract}

\end{titlepage}

\newpage


\setcounter{footnote}{0}
\section{Introduction}
\label{sec:intro}
\setcounter{equation}{0}

Recently Georgi \cite{Georgi,Georgi2} has suggested a new approach to analyzing the phenomenology of
conformal sectors coupled to the standard model by non-renormalizable interactions. He refers to the propagating degrees of freedom of such conformal sectors as ``unparticles" since their phase space resembles the phase space of a non-integer number of particles. Georgi \cite{Georgi} also speculated that it would be interesting to study the case where the unparticles had standard model gauge quantum numbers. Since the presence of new massless modes with standard model gauge couplings would drastically modify low energy phenomenology, we will assume in this paper that such charged unparticles have an infrared cutoff, so that the effects of the unparticles will be mainly restricted to high-energies such as will be probed by the Large Hadron Collider (LHC).  

The most spectacular signals for unparticles at the LHC would arise  in the case where the new unparticles carry a color charge.  For example we could imagine a theory with a new type of quark that carries color but also couples strongly to an approximate conformal field theory (CFT) sector. 
If sufficiently stable, these ``unquarks'' would appear in the detector as ``unjets'' which can be very different from the usual jets of QCD. Since, as mentioned above, we do not think there can be new light degrees of freedom that couple to gluons, the CFT must have some dynamics that cuts off the infrared at least below  the few hundred GeV range.  This is similar to the the case of Randall-Sundrum models \cite{RS} with gluons in the bulk: an infrared cutoff is required to have a phenomenologically acceptable  model, and the usual infrared cutoff (a TeV brane) violently breaks the approximate conformal symmetry of the model.  We will similarly impose a hard infrared cutoff on the colored unparticles, but it is still possible that color neutral unparticles can be massless or at least much lighter than the colored unparticles.

Essentially we are thinking of an effective theory for a massive particle that couples to a CFT so that it has a large anomalous dimension.  One of the intriguing features of Georgi's proposal is that it suggests new ways to impose an infrared cutoff on a CFT (or a Randall-Sundrum-like) theory.
 Further it may provide a weak coupling description of a CFT without relying on a large $N$ limit.
 
 Here we will focus on an analysis of colored scalar unparticle production from a quark-antiquark initial state.  Gluon fusion will probably dominate at the LHC, but that calculation is much more involved and we will be content here with demonstrating the general methods that go into such calculations. We will nevertheless provide a simple answer for the gluon fusion case as well. 
After introducing our infrared cutoff, we examine colored unparticle propagators and how to derive their gluon coupling Feynman vertices (form-factors). Finally we describe the parton level production cross-section calculation and finish off with conclusions and a discussion of future directions.

\section{An Infrared Cutoff}
\label{sec:cutoff}
\setcounter{equation}{0}

The phase space factor for a scalar unparticle final state with scaling dimension $d$ is \cite{Georgi}

\beq
d\Phi(p,d)=A_{d}\,\theta\left(p^0\right)
\,\theta\left(p^2\right)\,\left(p^2\right)^{d-2},
\label{georgiphasespace}
\eeq
where
\beq
A_d=\frac{16\pi^{5/2}}{(2\pi)^{2d}}
\,\frac{\Gamma(d+1/2)}{\Gamma(d-1)\,\Gamma(2d)}
\label{Ad}
\eeq
amounts to a useful normalization convention \cite{Georgi}.

Without knowing the details of how the scale invariance of the CFT sector is broken in the infrared, we will parameterize our ignorance with 
an infrared cutoff scale $m$. For a Lorentz scalar operator with scaling dimension\footnote{For scaling dimensions $d>2$ we need to provide an ultraviolet cutoff as well.} $1\le d<2$ we take the simple ansatz \cite{Fox:2007sy} :
\beq
\Delta(p,m,d)&\equiv& \int d^4 x \, e^{i px} \langle 0|T {\cal  O}(x)  {\cal O}^\dagger(0) |0\rangle 
\nonumber\\
&=&
 \frac{A_{d}}{2\pi}  \int_{m^2}^\infty (M^2-m^2)^{d-2} \frac{i }{p^2-M^2+i \epsilon}dM^2~.
 \label{massiveunparticle}
 \eeq
 This form has been chosen because in the limit $m\rightarrow 0$ it reduces to Georgi's unparticle
 propagator, and furthermore in the $d\rightarrow 1$ limit it reduces to a free particle propagator with mass $m$, which we know to be the case for gauge invariant operators from unitarity arguments \cite{unitarity}.  
 
 From the conjectured correspondence between five dimensional anti-de Sitter (AdS$_5$) space theories and CFTs, it also seems reasonable to suggest that the infrared cutoff  should not correspond to something like confinement since this would just result in towers of CFT ``mesons'', ``baryons'', and ``glueballs'', which are the Kaluza-Klein modes of the AdS$_5$ theory. Perhaps the CFT we are considering here has ``chiral symmetry breaking without confinement".  There are well understood CFTs where we can engineer such behavior.  Consider a Banks-Zaks fixed point theory \cite{BanksZaks} with a large $N$ gauge group. Giving a mass to one of the fermions will give an order $1/N$ correction to the fixed point, but the massive fermion will still be able to emit massless gauge bosons and will acquire a branch-cut in its propagator starting at the fermion mass.  By looking at supersymmetric examples we can extend this analysis to strong coupling.  Consider Seiberg's dual of SUSY QCD \cite{Seiberg}  in the conformal window at large $N$.  The meson operator has a scaling dimension between 1 and 2. We can assume for simplicity that the gauge coupling is tuned to its fixed point value in the UV. Adding a mass term for 
one of the flavors will again give an order $1/N$ correction to the fixed point and result in a propagator 
with a branch cut starting at the mass.  Such branch cuts are ubiquitous in field theory.  They occur in any theory where a particle can emit massless states such gauge bosons.  This happens for the electron propagator in QED \cite{Brown} and the gluon propagator in QCD \cite{Neubert}, for example.

 Performing the integration in (\ref{massiveunparticle}) we find
 \beq
\Delta(p,m,d)&=&
  \frac{A_{d}}{2 \sin d \pi} \frac{i}{\left( m^2-p^2-i \epsilon \right)^{2-d} }~.
   \label{scalarprop}
\eeq
In the limit $d\rightarrow 1$ we have 
\beq
\Delta(p,m,1)
&=&  \frac{i}{p^2-m^2+i \epsilon} ~,
\eeq
the free particle propagator, as required by unitarity \cite{unitarity}.

The discontinuity across the cut gives the modified phase space to be: 
\beq
d\Phi(p,m,d)=A_{d}\,\theta\left(p^0 \right)
\,\theta\left(p^2-m^2\right)\,\left( p^2-m^2\right)^{d-2} ~.
\label{phasespace}
\eeq
As $d\rightarrow 1$ from above, $d\Phi(p,m,d)$ approaches 1-particle phase space with the appropriate mass-shell constraint
\beq
d\Phi(p,m,1)=2 \pi \,\theta\left(p^0 \right)
\,\delta(p^2-m^2)~.
\eeq

\section{Gauge Interactions}
\label{sec:gaugeint}
\setcounter{equation}{0}

The propagator in Eq. (\ref{scalarprop}) can be derived from the momentum space  effective action
\beq
S=\frac{2 \sin d \pi}{A_d} \int \frac{d^4p}{(2\pi)^4}  \,\phi^\dagger(p) \left[m^2-p^2\right]^{2-d}
\phi(p) ~.
 \label{action}
\eeq
If we want to weakly gauge a global symmetry  $G$ of this theory we can first Fourier Transform the action (\ref{action}) to arrive at a non-local theory in position space 
\beq
S= \int d^4x \,d^4 y \,\phi^\dagger(y) F(x-y)
\phi(x) ~,
 \label{actionx}
\eeq
and then ensure $G$ gauge invariance by using Mandelstam's method \cite{Mandelstam} of introducing 
a path-ordered exponential of the gauge field,
 i.e. a Wilson line,
 \beq
 W(x,y)=P \exp \left[-ig T^a \int_x^y A^a_\mu dw^\mu\right]~,
 \eeq
  between the two unparticle fields evaluated at $x$ and $y$. This method has been previously applied to non-local toy-model field theories \cite{gauging,gauging2}, for example the non-local chiral-quark model  \cite{gauging2}. 
  
  Applying this method to the QCD interactions of the unparticles allows us to calculate the Feynman vertex for a gluon coupled to two unparticles.
The result using Eq.  (\ref{action}) is
\beq
i g \Gamma^{a\mu}(p,q)&\equiv&\frac{i\delta^3 S}{\delta A^{a\mu}(q)\delta \phi^\dagger(p+q)\delta \phi(p)} \nonumber\\
&=&
i g T^a \frac{2 \sin d \pi}{A_d}  \frac{2p^\mu+q^\mu}{2p \cdot q+q^2}\left[ \left(m^2-(p+q)^2\right)^{2-d}
-\left(m^2-p^2\right)^{2-d} \right]~.
\label{svertex}
\eeq
As a check we note that this vertex satisfies the Ward-Takahashi identity \cite{wt}
\beq
i q_\mu \Gamma^{a\mu} = \Delta^{-1}(p+q,m,d) T^a -T^a \Delta^{-1}(p,m,d)~.
\eeq
The path-ordered exponential includes arbitrarily high powers of the gauge field, so there are  vertices with arbitrary numbers of gauge bosons.  The  two gauge boson vertex is
\beq \label{svertex2}
g^2 \Gamma^{ab\mu\nu}(p,q_1,q_2) &=&- g^2  \left\{ \phantom{\frac{1}{2}}\left( T^a T^b + T^b T^a \right) g^{\mu\nu}\mathcal{F} (q_1+q_2)  \right. \\
&&\left.+ T^a T^b \frac{(2 p+q_2)^\nu (2 p + 2 q_2 + q_1)^\mu}{q_1^2 + 2 (p+q_2)\cdot q_1} \left[ \mathcal{F} (q_1+q_2) - \mathcal{F} (q_2) \right]  \right.\nonumber \\
&&\left.+ T^b T^a \frac{(2 p+q_1)^\mu (2 p + 2 q_1 + q_2)^\nu}{q_2^2 + 2 (p+q_1)\cdot q_2} \left[ \mathcal{F} (q_1+q_2) - \mathcal{F} (q_1) \right]\right\}\,,\nonumber
\eeq
where

\beq
\mathcal{F} (q) = \frac{2 \sin d \pi}{A_d} \,\frac{\left(m^2 - (p+q)^2\right)^{2-d}-\left(m^2-p^2\right)^{2-d}}{q^2 + 2 p\cdot q}\,.
\eeq

\section{Unparticle Production}
\label{sec:production}
\setcounter{equation}{0}

\begin{figure}[t]
  \centering
  \includegraphics[width=0.7\textwidth]{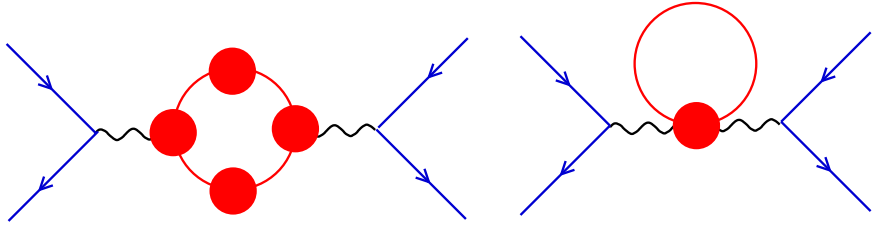}
  \caption{Forward scattering of $q{\bar q}$ through a gluon with a scalar unparticle vacuum polarization
  loop.  The imaginary parts of those 
  diagrams contribute to the unparticle production cross-section. }
  \label{fig:cut}
\end{figure}

In this section we want to compute the cross section $\sigma (q \bar q \to \mbox{unparticles})$ at leading order in $\alpha_s$. 
Thus we need to evaluate the imaginary part of $q\bar{q}$ forward scattering via a gluon in the $s$-channel with a one loop correction to the gluon line from unparticles, as shown in Fig.~\ref{fig:cut}.

A slick way to get the result is to calculate the effective action by taking the logarithm  of the partition function $Z$.
From writing (\ref{actionx}) as a power of a gauge covariant derivative $D$, we simply find from the path integral that:
\beq
\ln Z&=& -\frac{1}{2} \ln {\rm Det} (D^2+m^2)^{2-d}\nonumber \\
 &=&  -\frac{1}{2} {\rm Tr} \ln (D^2+m^2)^{2-d}\nonumber\\
  &=&  -\frac{1}{2}(2-d) {\rm Tr} \ln (D^2+m^2).
  \label{effectiveaction}
  \eeq
Thus the sum of one unparticle loop graphs with a fixed number of gauge boson legs is just $2-d$ times the usual one particle loop answer. 

We will verify this result by explicitly calculating the cross section in the $q\bar q$ channel. 
One might think of applying the usual Cutkosky rules~\cite{cutkosky}: cutting the unparticle propagators and replacing them with the unparticle phase space factor (\ref{phasespace}).
This however leads to the wrong answer, in particular a divergent cross section. The reason is that the vertices in Eqs.~(\ref{svertex}) and~(\ref{svertex2}) also contain non-trivial functions of the momenta, and they do contribute to the imaginary part of the $q\bar q$ scattering amplitude.
Therefore, we must consider the amplitude as a whole and compute the imaginary part directly. 

As a starting point, we can simplify the calculation by setting the infrared cut-off to zero. 
The imaginary part can be extracted from the vacuum polarization amplitude: for incoming momentum $q$, gauge invariance requires that such an amplitude is proportional to $q^2 ( \ln(-q^2/\Lambda^2) + \mbox{finite} )$, and the imaginary part can be extracted from the $\ln$ term 
\beq
{\rm Im}\left[ q^2 \ln(-q^2/\Lambda^2)\right]=q^2 \pi~.
\eeq
Since dimensional analysis requires that the $\ln(-q^2)$ comes with a logarithm of the regulator scale squared, 
$\Lambda^2$, we can easily find the coefficient of $\ln(-q^2)$ by extracting the regulator dependence.

The first diagram in Fig.~\ref{fig:cut} is proportional to:
\beq \label{eq:integral1}
\mathcal{A}_1 \sim \int \frac{d^4p}{(2 \pi)^4} \frac{q^2-4 p \cdot q+4 p^2}{\left(q^2-2p \cdot q\right)^2} 
\left[ 2-\left( \frac{-(p-q)^2- i \epsilon}{-p^2- i \epsilon}\right)^{2-d}-\left( \frac{-p^2- i \epsilon}{-(p-q)^2- i \epsilon}\right)^{2-d}\right]\,.
\eeq
Note that the normalization $A_d$ cancels between propagators and vertices in the loop as it should.
Looking at the large $p$ region of the loop integration we can Taylor expand for small $q$:
\beq
\int \frac{d^4p}{(2 \pi)^4} \frac{q^2-4 p \cdot q+4 p^2}{\left(q^2-2p \cdot q\right)^2} 
\left[ c_2   \frac{\left(q^2-2p \cdot q\right)^2} { p^4 } +c_3   \frac{\left(q^2-2p \cdot q\right)^3} { p^6 }  
+c_4   \frac{\left(q^2-2p \cdot q\right)^4} { p^8 } +\ldots   \right]\,,
\eeq
where
\beq
c_2&=&-(d-2)^2~,\nonumber\\
c_3&=&(d-2)^2~,\\
c_4&=&-(d-2)^2 ( d^2  - 4 d +15)/12~.\nonumber
\eeq
Picking out the terms that generate logarithms we are left with 
\beq
\int \frac{d^4p}{(2 \pi)^4} (c_2 +4 c_3)\frac{q^2}{p^4}+(c_3+2c_4) \frac{8 (p \cdot q)^2}{p^6}~.
\eeq
Thus, the first diagram's contribution to the  massless scalar unparticle production cross-section is suppressed relative to the
massless scalar particle production by
\beq \label{eq:suppression1}
\left. \frac{\sigma_d}{\sigma_1} \right|_{\mbox{diag. 1}} = \frac{d(2-d)^2(4-d)}{3}\,.
\eeq

The second diagram does not give any contribution in the particle case ($d=1$), where the only contribution to the imaginary part comes from cuts of propagators.
However, this is not true in the unparticle case.
The amplitude is proportional to the integral:

\beq \label{eq:integral2}
\mathcal{A}_2 &\sim&  \int \frac{d^4p}{(2 \pi)^4} \left\{ - \frac{2-d}{2} \frac{1}{p^2-m^2+i \epsilon} \left[ 8 +  \frac{q^2-4 p \cdot q+4 p^2}{\left(q^2-2p \cdot q\right)} + \frac{q^2+4 p \cdot q+4 p^2}{\left(q^2+2p \cdot q\right)}\right] \right.  \nonumber  \\
& +& \frac{q^2-4 p \cdot q+4 p^2}{\left(q^2-2p \cdot q\right)^2} \left[ 1-\frac{\left(m^2- (p-q)^2 -i \epsilon\right)^{2-d}}{\left(m^2-p^2-i \epsilon\right)^{2-d}} \right] \nonumber \\
&+& \left. \frac{q^2+4 p \cdot q+4 p^2}{\left(q^2+2p \cdot q\right)^2} \left[ 1-\frac{\left( m^2-(p+q)^2 -i \epsilon \right)^{2-d}}{\left(m^2- p^2 -i \epsilon \right)^{2-d}} \right]  \right\}\,.
\eeq
Applying the same technique as before, we find:

\beq \label{eq:suppression2}
\left. \frac{\sigma_d}{\sigma_1} \right|_{\mbox{diag. 2}} = \frac{(d-1) (d-2) (d^2 - 5 d +3)}{3}\,.
\eeq
As expected, this contribution to the cross section vanishes both in the $d\to 1$ and $d\to 2$ limits.
Combining the two results in Eqs.~(\ref{eq:suppression1}) and (\ref{eq:suppression2}) together, we obtain the simple result

\beq \label{eq:suppression}
 \frac{\sigma_d}{\sigma_1}  =  (2-d)\,,
\eeq
in agreement with Eq. (\ref{effectiveaction}).

In the Appendix we show how to obtain the same results via a direct calculation of the discontinuity of the integrals, providing a check of the results in Eqs~(\ref{eq:suppression1}) and~(\ref{eq:suppression2}).
Such a calculation  allows us to take into account the presence of the IR cutoff, which cannot be inferred from the small $q$ expansion above. 
Our simple ansatz for the  breaking of conformal invariance, leads to a very simple result: the $d$ and $m$ dependence factorize, and the dependence on the IR cutoff is given by the usual kinematical factor.
This result is achieved via non trivial cancellations between the two diagrams in Fig.~\ref{fig:cut}, as shown in the Appendix.
Therefore the unparticle production cross section is:

\beq
\sigma_d (m) = (2-d) \left(\sqrt{1-\frac{4 m^2}{q^2}}\right)^{3} \sigma_1 (m=0) = (2-d)\, \sigma_1 (m)\,.
\eeq

\section{Conclusions}

In this paper we have considered scalar unparticles that carry standard model gauge quantum numbers.
We have seen that the production cross section of such unparticles via gauge interactions is, at first order in $\alpha_s$, simply suppressed by a factor $2-d$ compared to the pair production of the corresponding scalar particles.
We explicitly calculated the cross section in the $q \bar q$ annihilation channel, where this simple result arises through a non trivial cancellation between two diagrams, one of which is does not contribute in the particle limit $d \rightarrow 1$.
We included a simple parameterization of an IR cutoff, which seems phenomenologically necessary for  unparticles to be charged under standard model gauge groups.
Our parameterization yields a simple result: the IR cutoff dependence factorizes and the cross section is multiplied by a  simple  suppression factor.
These results may be generalized to the gluon fusion channel, which will dominate for unparticle pair production at the LHC.
It would be quite interesting to study the detailed phenomenology of colored unparticles at the LHC.

In addition to having a suppressed cross-section, the final states arising from these colored unparticles would look very different from ordinary jets. If the unparticles are stable enough to travel a significant distance through the detector one would have to understand the details of unparticle QCD hadronization (``unjet'' formation). Some of the difficulties involved are analogous to those of  the long-lived gluino scenario \cite{gluinos}. A further complication is that the unquarks can still radiate into lighter CFT degrees of freedom that are not colored (the gauge singlet  unparticles usually considered in the literature). Thus there is an additional loss of energy that is invisible to the detector, which results in missing energy along the two unjet directions.
In a more general model than we have considered here, with mixing between colored unparticles and other colored particles, one could imagine a scenario where a single colored unparticle is produced so that the missing energy is entirely along a single unjet direction.

An even more interesting scenario to explore is one where the unparticles also have electroweak quantum numbers.
Unquarks could then decay into ordinary quarks and they would have a passing resemblance to a fourth family \cite{fourth}.  (This is the simplest way to allow unquarks to decay and avoid searches bounding  stable exotics.)
We could even imagine that the unquarks play a role in electroweak symmetry breaking which would give us a new approach to the hierarchy problem. If the scaling dimension of the analog of the Higgs mass operator is larger than two, such a scenario would have some similarities with gaugephobic Higgs models \cite{gaugephobic}.

\section*{Acknowledgments}
We thank Andy Cohen, Csaba Cs\'aki, Howie Haber, Tao Han, Mathias Neubert, Matt Reece, Pierre Ramond, Jose Santiago, Yuri Shirman,  and  Arkady Vainstein for useful discussions and comments.
The authors are supported by the US department of Energy under contract DE-FG02-91ER406746. 
GM and JT  thank the Aspen Center for Physics where part of this work was completed.

\appendix

\section{Appendix: Calculation}
\label{app:calculation}
\setcounter{equation}{0}

In this appendix we discuss an alternative way to calculate the integrals in Eqs~(\ref{eq:integral1}) and (\ref{eq:integral2}), that correctly takes into account the effect of the IR cutoff of the conformal sector.

Let us focus on Eq.~(\ref{eq:integral1}):

\beq \label{eq:integral1App}
\int \frac{d^4p}{(2 \pi)^4} \frac{q^2-4 p \cdot q+4 p^2}{\left(q^2-2p \cdot q\right)^2} 
\left[ 2-\left( \frac{m^2-p^2-i \epsilon}{m^2-(p-q)^2-i \epsilon}\right)^{2-d}-\left( \frac{m^2- (p-q)^2-i \epsilon}{m^2 - p^2-i \epsilon}\right)^{2-d}\right].
\eeq
We are interested in the imaginary part of (\ref{eq:integral1App}), so we can neglect the factor 2 inside the square brackets. 
For simplicity, we will illustrate the calculation on the second term in the integral:

\beq
\Xi (q) = \int \frac{d^4p}{(2 \pi)^4} \frac{q^2-4 p \cdot q+4 p^2}{\left(q^2-2p \cdot q\right)^2}  \left( \frac{m^2-p^2-i \epsilon}{m^2-(p-q)^2-i \epsilon}\right)^{2-d}\,.
\eeq

We use Feynman parameters to rewrite this integral as
\beq
{\rm Im} \; \Xi(q) = - \frac{\sin d \pi}{\pi} \; {\rm Im} \int \frac{d^4p}{(2 \pi)^4} \frac{q^2-4 p \cdot q+4 p^2}{\left(q^2-2p \cdot q\right)^2} \int_0^1 dx \left[  \frac{x^{1-d} (1-x)^{d-2} \, (p^2 - m^2)}{p^2+x (q^2 -2 p \cdot q) -m^2+i \epsilon} \right]\,.
\eeq
In this way the $d-$dependence appears only in the powers of $x$: momenta all have integer powers. Going to the rest frame, where $p=(E,{\bf \overrightarrow{p}}), q=(q, \overrightarrow{0})$, we notice that the integrand has a pole at $E=q x+\sqrt{p^2-q^2 (1-x) x}$. Taking the residue at this pole we get

\beq \label{eq:imag2}
{\rm Im} \; \Xi(q) &=& - \frac{\sin d \pi}{8 \pi^3}  \;{\rm Im} \int_0^{\infty} dp \int_0^1 dx \left(\frac{x}{1-x}\right)^{2-d} \nonumber \\
 & & \cdot \; \frac{p^2 \left(4 m^2+q \left(4 \sqrt{m^2+p^2-q^2 (1-x) x} (2 x-1)+q (1-8 (1-x) x)\right)\right)}{q \sqrt{m^2+p^2-q^2 (1-x) x}
   \left(q (2 x-1)+2 \sqrt{m^2+p^2-q^2 (1-x) x}\right)} \;\;\;\;\;\;\;
\eeq
Next notice that the term $(q (2 x-1)+2 \sqrt{m^2+p^2-q^2 (1-x) x})$ appearing at the denominator of Eq. \eq{imag2} vanishes at $p=\sqrt{q^2-4 m^2}/2$ only if $x<1/2$. Thus this pole will give a contribution to the imaginary part of the integral. We evaluate it taking the residue
\beq \label{eq:imag3}
{\rm Im_{pole}} \; \Xi(q) &=&  \frac{q^2}{32 \pi^2} \sin d \pi  \; \beta^{3}  \int_0^{1/2} dx\; \left(\frac{x}{1-x}\right)^{2-d} \nonumber  \\
& = &  \frac{q^2}{32 \pi^2} \sin d \pi \; \beta^{3}\; B\left(\frac{1}{2}; 3 - d, -1 + d\right) 
\eeq
where $\beta =\sqrt{1-4 m^2/q^2}$ and $B$ is the incomplete beta function, defined as

\beq
B\left( z; a, b\right) = \int_0^z du\; (u)^{a-1} (1-u)^{b-1}\,.
\eeq

There is another contribution to the imaginary part of our original integral, which comes directly from the region where the argument of the square roots appearing in Eq. \eq{imag2} is negative. This can be evaluated as
\beq 
{\rm Im_{\sqrt{}}} \; \Xi(q) &=& - \frac{\sin d \pi}{8 \pi^3}   \int_{(1-\beta)/2}^{(1+\beta)/2} dx\;  \left(\frac{x}{1-x} \right)^{2-d}   \nonumber \\ 
& & \cdot \int_0^{\sqrt{-m^2+q^2 x(1-x)}} \hspace{0cm} dp  \;
 \frac{p^2 \left(4 m^2+8 p^2-q^2\right) (1-2 x)}{\left(4 \left(m^2+p^2\right) -  q^2\right) \sqrt{-m^2-p^2+q^2 (1-x) x}}
\eeq
The integral can be performed analytically and the result can be expressed in terms of incomplete Beta functions:

\beq \label{eq:imag4}
{\rm Im_{\sqrt{}}} \; \Xi(q) & =&   \frac{q^2}{32 \pi^2}  \sin d \pi \left[ - \beta^3 \; B \left(\frac{1}{2};3-d,d-1 \right)  + \beta^3 \; B(\frac{1-\beta }{2} ; 3-d,d-1)+ \;\;\; \right. \nonumber \\ 
&+&  \frac{\beta^3 + 2 \beta^2 - 1}{2} \; f(\beta, 3-d, d-1) + (3-2 \beta^2) \; f(\beta, 4-d, d-1)  + \nonumber  \\
& -&\left. \phantom{\frac{1}{2}}  \hspace{-0.38cm}  6 \; f(\beta, 5-d, d-1) + 4 \; f(\beta, 6-d, d-1) \right]   
\eeq
where we defined the auxiliary function
\beq
f(\beta,a,b) &=& B_{}\left(\frac{1+\beta }{2};a,b\right) -B_{}\left(\frac{1-\beta }{2};a,b\right)\,.
\eeq
It is interesting to note that the contribution of the last term in Eq.~\ref{eq:integral1App} can be obtained by the previous result substituting $d \to 4-d$.

The contribution to the cross section of the 2 diagrams in Fig.~\ref{fig:cut} can be written as:

\beq
\left. \frac{\sigma_d (m)}{\sigma_1} \right|_{\mbox{diag. 1}} &=& \frac{\sin d \pi}{\pi}  \left[\frac{\beta^3 + 2 \beta^2 -1}{2} \Big( f(\beta, 3-d, d-1) - f(\beta, d-1, 3-d) \Big) +\right.  \nonumber \\
& + & \beta^3  \Big( B(\frac{1-\beta }{2} ; 3-d,d-1) - B(\frac{1-\beta }{2} ; d-1,3-d) \Big) +  \nonumber  \\
& +& (3 - 2 \beta^2 ) \Big( f(\beta, 4-d, d-1) - f(\beta, d, 3-d) \Big) +  \nonumber \\
&-& 6 \Big( f(\beta, 5-d, d-1) - f(\beta, d+1, 3-d) \Big) +  \nonumber \\
&+& \left. \phantom{\frac{1}{2}} \hspace{-0.38cm} 4 \Big( f(\beta, 6-d, d-1) - f(\beta, d+2, 3-d) \Big) \right]\,;
\eeq

\beq
\left. \frac{\sigma_d (m)}{\sigma_1} \right|_{\mbox{diag. 2}} = (2-d) \beta^3 - \left. \frac{\sigma_d (m)}{\sigma_1} \right|_{\mbox{diag. 1}}\,.
\eeq
where the result from the second diagram matches non trivially the same series of Beta functions as the first diagram, leading to the simple result presented in Section~\ref{sec:production}.

We can now check that in the case $m\to 0$ we reproduce the results in Eqs~(\ref{eq:suppression1}) and (\ref{eq:suppression2}), using the following properties of the Beta functions:
\beq
B \left(0; a,b\right) = 0\,; \quad B(1;a,b)=\frac{\Gamma(a) \Gamma(b)}{\Gamma(a+b)}\,, \qquad   \mbox{if} \quad a,b > 0\,.
\eeq

\end{document}